# Revealing the chemical bonding in adatoms arrays via machine learning of 3D scanning tunneling spectroscopy data

Kevin M. Roccapriore[1], Qiang Zou[1], Lizhi Zhang[2], Rui Xue[3], Jiaqiang Yan[4], Maxim Ziatdinov[1], Mingming Fu[1], David Mandrus[3], Mina Yoon[1], Bobby Sumpter[1], Zheng Gai[1*] and Sergei V. Kalinin[1*]

[1] Center for Nanophase Materials Sciences, Oak Ridge National Laboratory, Oak Ridge, TN 37831

[2] Department of Physics and Astronomy, University of Tennessee, Knoxville, TN 37996, USA

[3] Department of Materials Science and Engineering, University of Tennessee, Knoxville, TN 37996, USA

[4] Materials Science and Technology Division, Oak Ridge National Laboratory, Oak Ridge, TN 37831

The adatom arrays on surfaces offer an ideal playground to explore the mechanisms of chemical bonding via changes in the local electronic tunneling spectra. While this information is readily available in hyperspectral scanning tunneling spectroscopy data, its analysis has been considerably impeded by a lack of suitable analytical tools. Here we develop a machine learning based workflow combining supervised feature identification in the spatial domain and un-supervised clustering in the energy domain to reveal the details of structure-dependent changes of the electronic structure in adatom arrays on the $Co_3Sn_2S_2$ cleaved surface. This approach, in combination with first-principles calculations, provides insight for using artificial neural networks to detect adatoms and classifies each based on their local neighborhood comprised of other adatoms. These structurally classified adatoms are further spectrally deconvolved. The unexpected inhomogeneity of electronic structures among adatoms in similar configurations is unveiled using this method, suggesting there is not a single atomic species of adatoms, but rather multiple types of adatoms on the $Co_3Sn_2S_2$ surface. This is further supported by a slight contrast difference in the images (or slight size variation) of the topography of the adatoms.

The functionality and structure of a material are intrinsically connected to the nature of the chemical bonding between the constituent elements,[1,2] which is deep-rooted within the field of chemical graph theory[3] and even critical to even the pharmaceutical research industry.[4] Correspondingly, the drive towards understanding the bonding mechanisms in solids, both to achieve a greater understanding of materials and enable property predictions, have been a hallmark of materials science and condensed matter physics throughout the last century. Traditionally, this knowledge was obtained using a combination of progressively more advanced theory, macroscopic measurements, and scattering techniques that shed light both into the structure of materials, as well as the minute details of electronic density distributions. In particular, recent advances in ultrafast scattering methods[5,6] have enabled insight into the very mechanisms of chemical bond formations. However, most scattering studies by their very nature are limited to macroscopically averaged properties and hence necessitate either translational symmetry, or sampling of a large number of identical objects.

In this regard, of special interest are the surfaces of materials. Surfaces are the active players in catalysis, heterogeneous reactivity, and electrochemical processes, and has naturally evolved into a field of its own. For several decades, surface studies were underpinned by spectroscopic techniques, low-energy electron diffraction, and more recently surface-sensitive X-ray methods. However, it was the development of atomically resolved scanning probe microscopy (SPM) techniques, notably scanning tunneling microscopy (STM)[7] and non-contact atomic force microscopy (nc-AFM)[8] that enabled direct visualization and probing of surfaces – even insulators – on the atomic level. Consequently, STM studies of metals, semiconductors, and other functional materials are now mature techniques that have presented new insights into surface structure and electronic properties and have enabled observations of chemical and physical processes.

However, despite the remarkable progress in STM and nc-AFM and the availability of large volumes of high-quality data, the potential of these techniques to gain insight into the fundamental mechanisms of surface behaviors has remained limited. This limitation stems from a dearth of algorithmic tools that can be used to extract relevant details from hyperspectral data sets. In fact, the vast majority of these STM and nc-AFM studies are performed using either single point spectroscopy, linear spectral unmixing techniques such as principal component analysis, or exploration of the energy-dependent Fourier space of the system.

Here we develop a machine learning (ML) workflow combining supervised feature identification in the spatial domain to identify atomic configurations, and non-supervised clustering in the energy domain to reveal structure-dependent changes of the electronic structure.

Specifically, this approach is used to reveal details of the electronic structure in sulphur adatom configurations on the Co-Sn cleaved surface of $Co_3Sn_2S_2$. The variability of electronic behaviors due to the local atomic neighborhoods provides insight into the electronic structure changes caused by the formation of lateral chemical bonds.

The ball and stick model of the $Co_3Sn_2S_2$ crystal is shown in Figure 1 (a). Two sulphur layers and one tin layer are sandwiched by a kagome layer consisting of Co and Sn atoms (Co/Sn layer). During the low-temperature cleavage process, several types of surface terminations can be produced as reported previously[11–15]. Shown in Figure 1 (b) is a STM topography image of one surface morphology we found other than the previously reported ones. A large number of adatoms, most of which cluster in the form of a chain, dimer, or monomer, are scattered on the hexagonal lattice (Co/Sn layer) surface. A line profile was acquired along the red line in Fig. 1 (b) and shown in Fig. 1 (c); the height of the atoms is around 0.16 – 0.17 nm, with the edge atoms, e.g., dimers, monomers and end atoms of the chains, slightly higher (0.01 nm) than the rest of the adatoms, which suggests the adatoms are S adatoms. Several adatoms (four atoms circled in blue in Fig. 1 (c)) - have a height of 0.22 nm (see line profile Fig. 1 (d) along blue line), consistent with the height of Sn adatoms. The drastic difference in spectra from the subsurface Co/Sn layer and the average of a chain of adatoms shown in Fig. 1 (e) confirms the different chemical compositions of those features. It is interesting to note that small but visible differences exist among the S adatoms within the chain that depend on their neighboring adatoms.

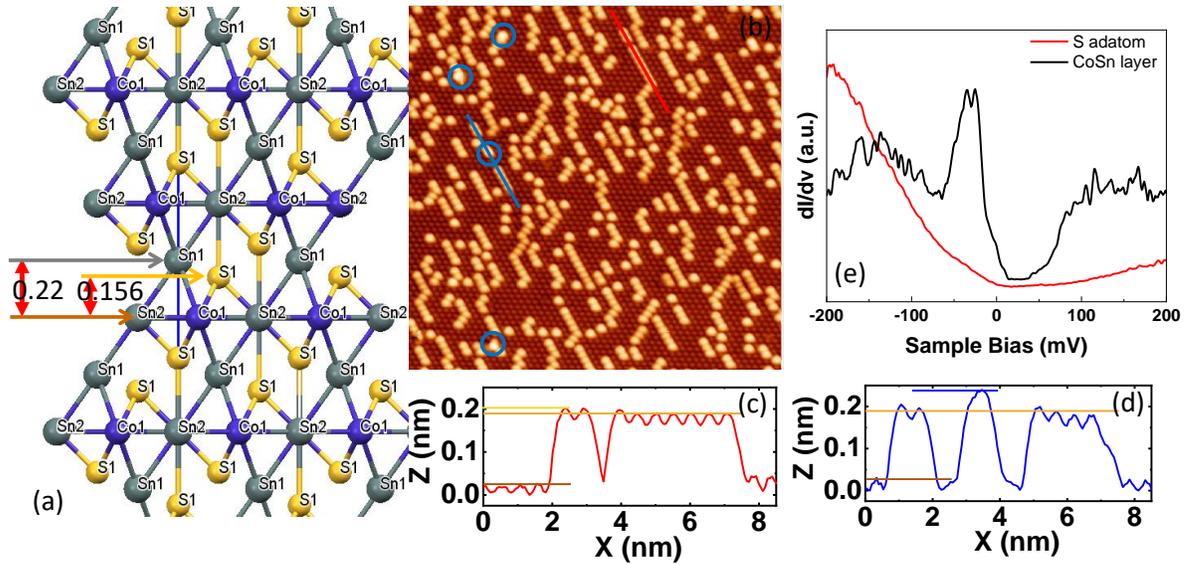

**Figure 1**. (a) Ball and stick model of $Co_3Sn_2S_2$ crystal. (b) STM image of one of cleaved surfaces of $Co_3Sn_2S_2$ (30 nm x 30 nm, 0.1 nA, -200 mV)), adatoms circled in blue are higher than rest of the atoms. (c) Line profile along red line in (b) shows height of one adatom chain is 0.16 - 0.17 nm with slightly higher edge atoms. (d) Line profile along blue line in (b) shows height of a separate individual adatom within a chain is 0.22 nm, considerably higher than the previous adatom chain, suggesting Sn atomic species. (e) Spectra of Co/Sn subsurface (black), and average of S adatom chain (red).

To understand the electronic changes induced by adatom bonding, we explored how the tunneling spectra of adatoms and consequently their electronic density of states (DOS) depended on the structural configuration relative to the neighboring adatoms on a surface. Since the variability of the DOS at the Fermi level can, in the first approximation, be attributed to the local structure, only the nearest neighbors should contribute appreciably to the modification of this behavior. To gain better insight, we derived the statistics of the local electronic DOS obtained by scanning tunneling spectroscopy (STS) that correspond to the different adatom configurations, and to this end, we employed a ML-based approach to reveal the relationship between the DOS and adatom configurations. Previous studies of the same $Co_3Sn_2S_2$ material by Morali et. al.[12] and Hasan[13] showed different surface terminations resulting from different cleavage planes, but the effect of the active local adatoms and their consequence on global material surface properties was not discussed, e.g., the topology, which is a primary reason for exploring this type of a Weyl semimetal.

To gain initial insight into the spatial variability of the electronic properties of the surface, we attempted the classical multivariate statistical analysis of the spectral data cube. This approach was pioneered for electron energy loss spectroscopy (EELS) in 2006 by Bosman[16] and for scanning probe microscopy (SPM) in 2009[17] and is now standard in these fields.[18,19] Generally, in the unsupervised linear unmixing methods, the 3D data cube containing the tunneling spectra at each spatial location on the grid is represented as a linear combination of the endmembers, where $Iv_i(V)$ represents the characteristic behaviors, and loading maps, $a_i(x,y)$, represent the spatial variability of these behaviors, as:

$$\frac{dI}{dV}(x,y,V) = \sum_{i=1}^{N} Iv_i(V)\, a_i(x,y) \qquad (1)$$

In this manner, the dimensionality of the data set is significantly reduced, corresponding to transition from $MN^2$ points for the full 3D data set on an $N \times N$ spatial grid and M energy points to K endmembers of M point each and K loading maps of size $N \times N$.

Both the endmembers and the weights can be optimized using specific constraints on non-negativity, additivity, sparsity, etc. For example, classical principal component cnalysis (PCA) performs the decomposition assuming orthogonality of the endmembers and ordering them in the direction of reduced variance, but does not impose any specific constraints on sparsity or additivity of the endmembers or loading maps. Non-negative matrix factorization imposes non-negativity condition on end members, whereas independent component analysis attempts to minimize the Gaussianity during the decomposition. Other constraints can be defined as guided by the physics of the problem.

The endmembers and loading maps of the first six PCA components, along with the associated fast Fourier transforms (FFTs) are shown in Fig. 2. The FFTs of the 1$^{st}$ and 3$^{rd}$ components clearly show peaks corresponding to the surface lattice, while the corresponding maps show the variations corresponding to adatom geometries. However, identification of the adatom ordering in the FFT can be speculative at best due to the low coverage (0.18) and high disorder in the adatom sublattice. The contrast variations corresponding to individual adatoms are detected in several of the PCA components, namely 3 and 5, with varying strengths for different components. Note that the PCA analysis seems to suggest different properties for different adatoms as manifested by different contrast, but fails to spatially resolve or separate these effects. This analysis, as well as exploration using alternative linear unmixing methods such as non-negative matrix factorization (NMF) and independent component analysis (ICA) which are examined in the available Jupyter notebook, generally demonstrates that there is a variation of electronic properties between adatoms and the surface; however, the stochasticity in the electronic properties of the surface does not allow clear identification of the behavior characteristic for adatom groups depending on bonding.

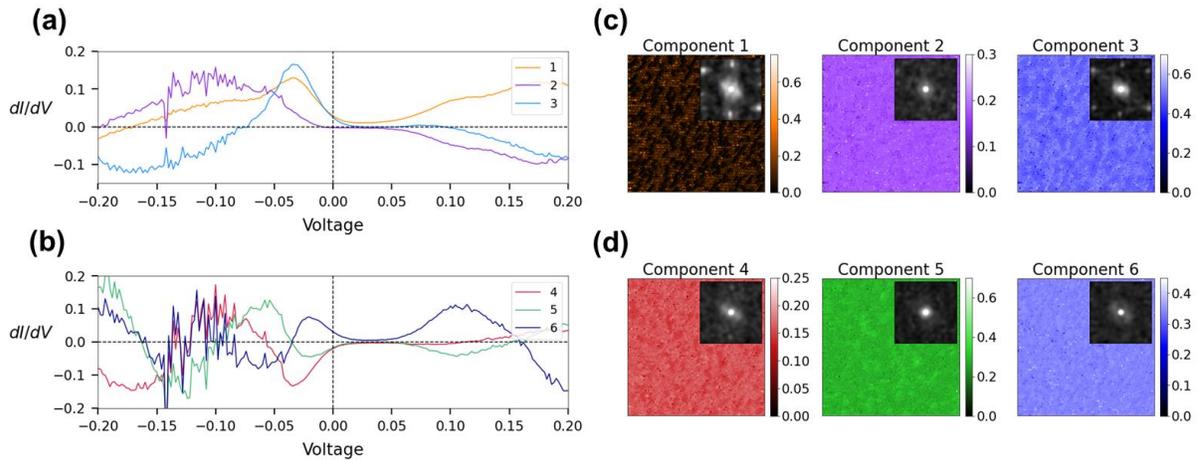

**Figure 2**. (PCA) of first six statistically relevant components. Decomposed spectral components are shown in panels (*a*) and (*b*) while corresponding loading maps are shown in (*c*) and (*d*). Insets are FFTs of the loading maps, which provide an indication of the unmixed feature's periodicity. Note the spectral components are separated in two plots only for clarity.

This consideration necessitates using the feature-oriented ML approach, where the specific classes of adatoms are identified based on their nearest neighbors, and the electronic properties are compared between these groups. The workflow for such an analysis is developed as follows. First, the inplanar atomic coordinates of adatoms are found using the deep convolutional neural network[20,21], adapted for this dataset from the earlier developed U-Net[22]. Using these coordinates, each adatom is classified into a structural configuration class based on possible adatom lattice configuration (supervised classification step). After sorting each adatom into different structural classes, the tunneling spectra for all adatoms within each class are statistically analyzed to retrieve their mean and standard deviations, as depicted in Figure 3.

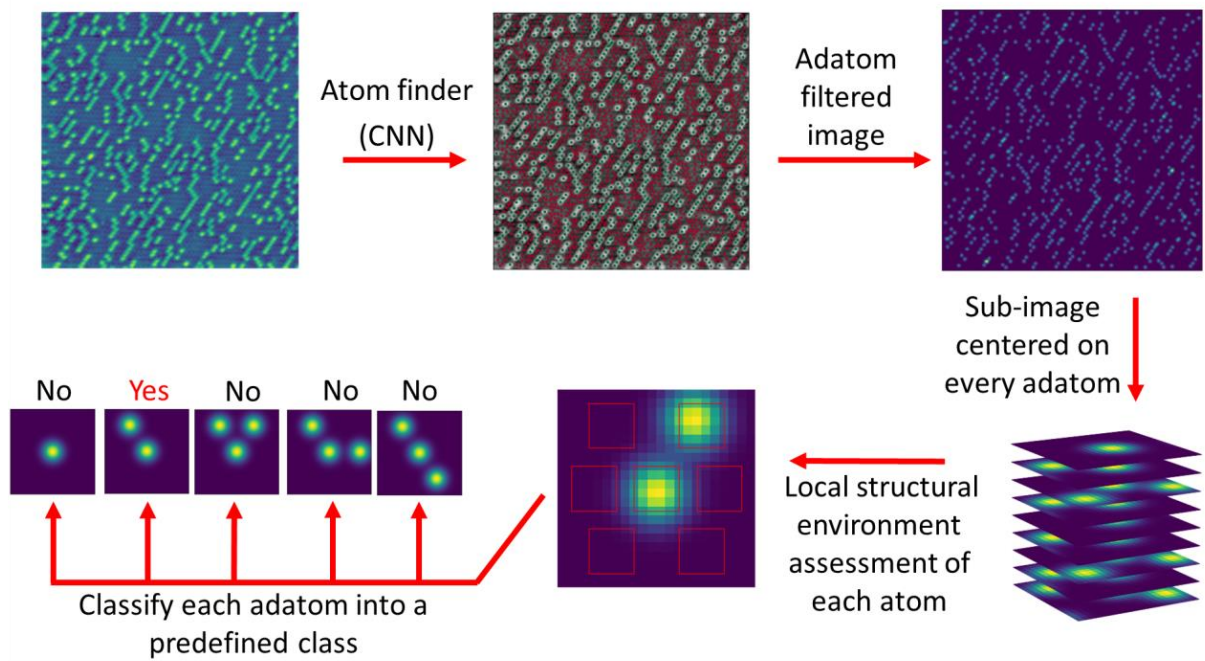

**Figure 3.** Analysis workflow comprised of atom identification and supervised classification. Adatom coordinates are determined using a convolutional neural network. At each ad-atom site, a smaller "sub-image" is made showing local structural environment up to first nearest neighbors of the feature and this local environment is analyzed in specific regions of interest to classify into predetermined structural configurations. Finally, spectra within each class are statistically analyzed.

To identify atomic positions, a convolutional neural network trained on synthetically generated images of a hexagonal lattice populated with various adatoms randomly distributed across the surface was used to locate adatoms and their coordinates in the experimental data. The network is built from three levels: first, a set of convolutional layers activated by the rectified linear unit (ReLU) function; second, a max-pooling level, and third, an up-sampling block. This architecture is chosen due to its relative simplicity and is commensurate with the level of feature detection required here.

For training, a set of synthetic images resembling the experimentally acquired data are created. The lattice rotation, periodicity, pixel density, and sizes of both types of atoms were varied to allow for robust semantic segmentation. Starting from coordinates of a perfect hexagonal lattice, background atoms in the form of 2D Gaussians are placed at small, random fluctuations away from these coordinates, to imitate a slightly imperfect lattice, which is typically observed due to distortions and instabilities in scanning-based systems. Next, a portion of the atoms from random locations are removed and replaced with larger and brighter atoms to mimic the sulfur adatoms. A sliding window is swept across the entire image to generate subsets of images in order to have a larger training sample size. To further generalize the training data, each subset had a random amount of zoom and Poisson noise applied to it. These steps were repeated several times to generate a large training dataset with a generous amount of variability to represent the expected differences among experimentally collected data. Once trained, the network is used primarily to detect the adatom spatial coordinates, although surface atoms can be detected as well.

It should be noted that prior to the deployment of the network, simpler approaches using standard blob detection directly on the experimental data via calculation of Laplacian of Gaussian (LoG) were attempted. However, LoG was less effective at atom finding due to the variations in adatom size and intensity depending on the adatom neighbourhood. Atom finding with the neural network was of additional importance compared to other methods and has the advantage of increased speed, which is attractive in our case where many statistics are desired. Since the network is trained on a variety of atom sizes and intensities, it only needs to be trained once and can subsequently identify a variety of different atom intensities and sizes among a range of datasets. Other methods require human input for each new dataset, at an increased cost of time, that can also introduce a source of possible bias that should be avoided. Equally important is that atom finding can be performed using less sophisticated methods if a higher pixel to atom ratio were present; however, due to the nature of hyperspectral data, longer acquisition times per pixel are required and consequently, a fewer number of total pixels can be reasonably acquired without considering stability issues (e.g., drift).

With the atomic segmentation accomplished and adatom positions identified, the type of local neighborhood for each adatom based on the number and orientation of nearest neighbors is identified. As a first step, the feasibility of the unsupervised classification of the adatom geometries was explored. Here, a set of 2D sub-images centered at each recognized adatom was created, and the sub-image stack was analyzed using the Gaussian mixture model (GMM), as illustrated in Figure 4. The GMM allows for separation of characteristic configurations, but it does not allow for reconstruction of expected pure configurational classes in agreement with prior work.[23] After much experimentation, this behavior was traced to intrinsic variability among atoms in different bonding states, which directly affects the electronic DOS. For example, if only a small number of atoms exist in, e.g., a trimer orientation in which neighbors are separated by 60°, GMM prioritizes rotations of the same orientation that appear more frequently. With the increased number of GMM components, the physically real class will start to be separated in subclasses prior to an unphysical class separated into physics. Similar issues emerge with different clustering methods. Most importantly, physically relevant configurations should account for rotations and mirror symmetries of the point group, whereas GMM classes do not.

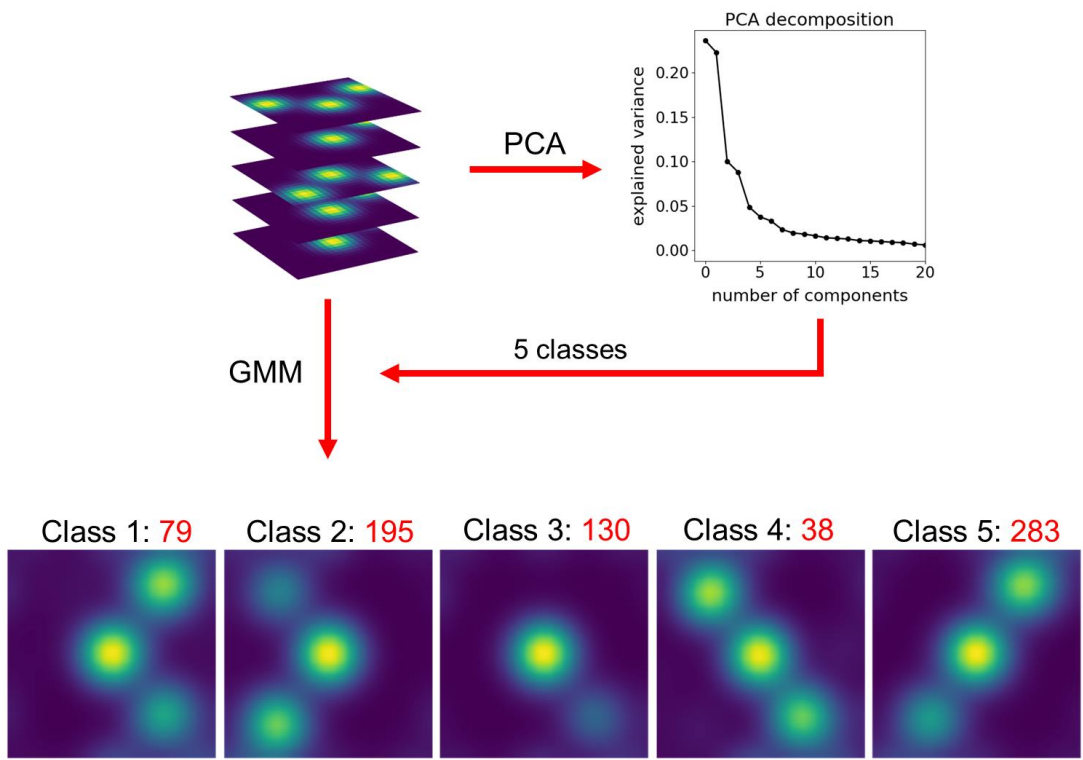

**Figure 4.** GMM used as a classifier to separate structural configurations in an unsupervised manner. PCA of the image stack informs on the number of classes to be used with GMM clustering. Note that these unsupervised classes are unphysical, i.e., the intensities of peripheral adatoms can be significantly below that of central atom.

To avoid this problem, we employed a supervised classification approach, where intensity thresholding is used to determine if an atom is present in one of six possible predetermined regions of interest (ROI). These ROIs are chosen based on the possible adatom positions consistent with the orientation of the hexagonal lattice in the dataset. In the general case of arbitrary rotation of the lattice, an FFT-based approach could be used to realize the appropriate location of each ROI as well as the rotation of each class designation, or alternatively, the data can be horizontally aligned by applying a rotation matrix and the same ROIs can be used. Based on the combination of binary output of each ROI's threshold, each sub-image can be categorized into one of the defined classes. Because the spectral data is already aligned to the structural data, the only step that remains is to organize the spectra of each class and statistically analyze it. The simplest approach is to compute the average spectrum of all the spectra within each class, as well as a standard deviation, the result of which is depicted in Figure 5. The clustering of the classes in the energy domain in the PCA space formed by the two dominant eigenvectors is also shown (Figure 5(b)).

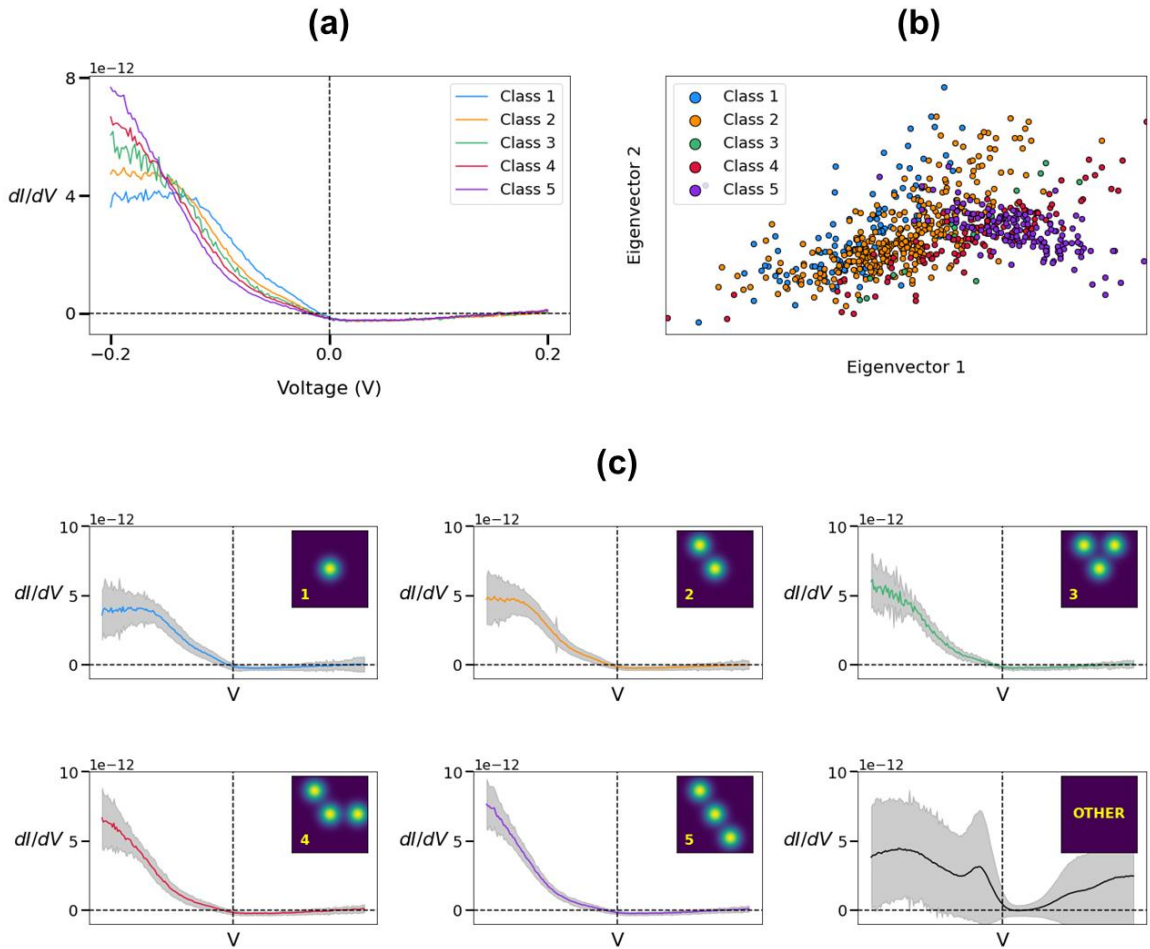

**Figure 5.** (*a*) Average spectra of each class overlaid simultaneously; (b) visualization of clustering of classes in PCA space, specifically clustering in space defined by first two eigenvectors; (c) each class average and standard deviation shown separately, where shaded black region is standard deviation. Insets show the atomic configuration for class number. Note all vertical axes have the same range, with voltage sweep from -200 mV to +200 mV in all cases.

From inspection of the spectral classification, it is clear from the average spectra that there is a significant difference among each of the five physical classes. The sixth panel denoted as "other" is simply the average and standard deviation of all other pixels not identified and is actually the majority of the (x,y) data points; hence, the standard deviation is expected to be relatively large. We emphasize that this separation was not possible with standard unsupervised decomposition techniques (e.g., GMM, PCA), but requires supervision to extract relevant information.

This analysis can be further extended to gain new information regarding the chemical bonding states of adatoms and their relation to the electronic structure. Upon examination of the large distribution of electronic behaviors in the PCA space as shown in **Figure 5 (b)**, we note that class 2 contains the greatest number of atoms and has the largest standard deviation. Figure S3 shows additional aspects and representations of clustering in both two and three dimensional PCA space. The large spread of the clusters in PCA space suggests that there may be additional mechanisms that give rise to the electronic inhomogeneity on the material surface. This prompts us to explore additional unmixing within each structural class.

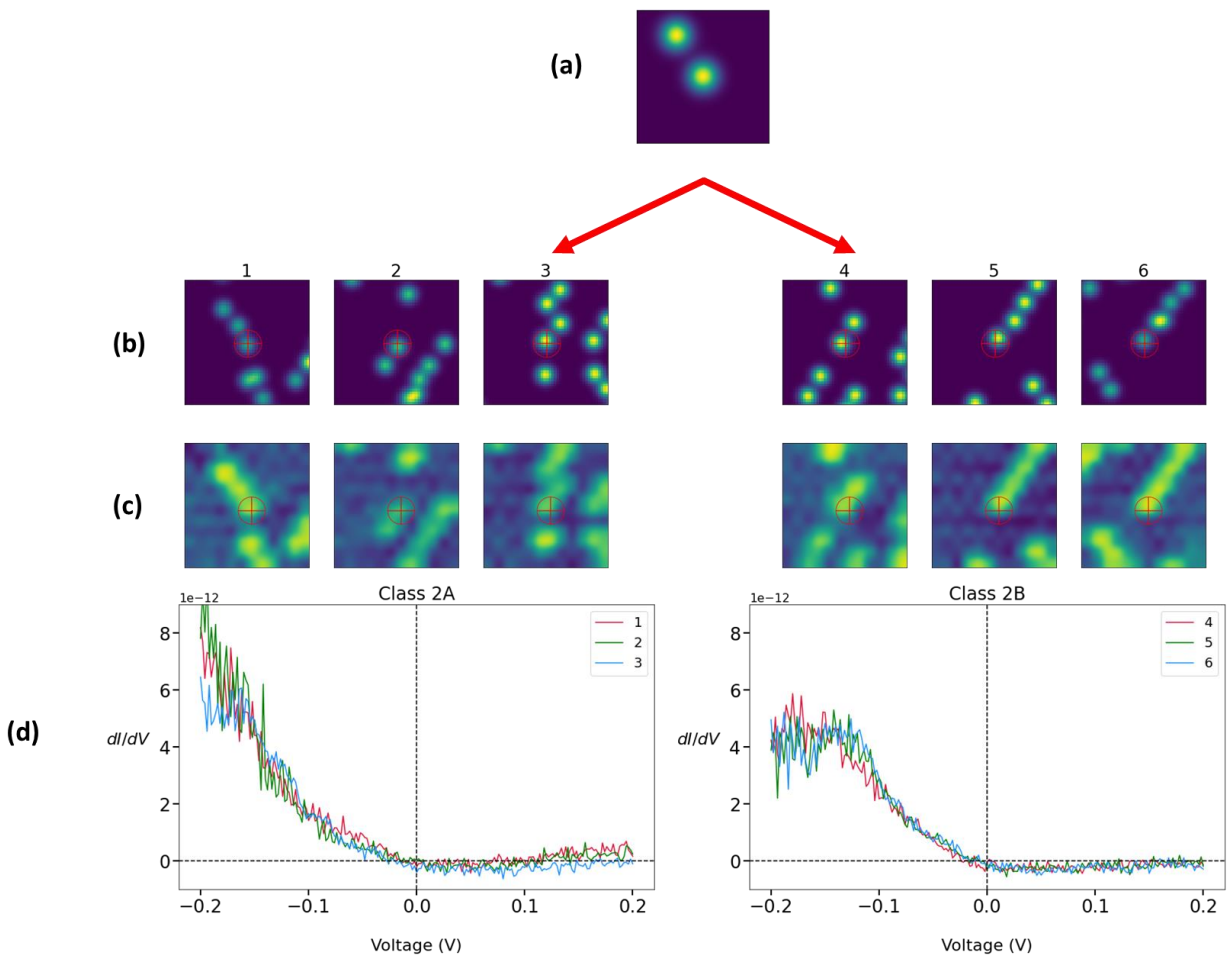

**Figure 6**: Selected atoms after refinement of class 2 by PCA decomposition filtering. (a) Depicts starting structural class, while panels (b) and (c) show cropped windows for selected subclassified atoms in adatom filtered image and upscaled topography, respectively. Corresponding spectra for center atoms in (b) and (c), highlighted by a crosshair, are shown in (d). Note all six adatoms have a single nearest neighbor.

Attempts were made to refine this classification further by relying on an unsupervised and purely spectral decomposition, as illustrated in **Figure 6**. Using the second component from the PCA decomposition as the class separator (note the first component relates to the average while the second is associated with the next most dominant behavior), a threshold value is employed such that if an adatom's PCA component strength is greater than this threshold, it becomes part of the new subclass. **Figure 6** demonstrates that the structural classification was successful by showing several of the local environments of the two sub-classes of class 2; each atom within both sub-classes of class 2 has only a single nearest neighbor, yet the electronic properties measured via dI/dV are substantially distinct. As shown in Figure S4, we were able to separate the clusters of class 2 in PCA space reasonably well using this method, and in PCA space it is clear there are two distinct classes that have not been artificially separated. Note that adjusting probe position by a pixel in any direction does not change this outcome. Despite all atoms within the original class 2 having been properly classified *structurally* as per our specifications, there is a striking spectral difference. The average of the class spectra with the new separation of classes is shown in **Figure S5.** In any case, there is a clear structural similarity among these two new sub-classes in the nearest neighbor limit, while exhibiting dissimilar spectra.

This leads to two possible outcomes: either (a) atoms beyond the first nearest neighbors have an impact on the electronic DOS of the adatom or (b) different atomic species of adatoms are observed. We also considered observation of subsurface effects but we believe the other possibilities suggested are more likely, while subsurface probing is difficult to confirm. After clustering of the subclasses using GMM methods (Supplemental Figure S6), no correlation between second or further nearest neighbors and the observed spectral behavior is observed, i.e., the local DOS of the adatoms depends only on the local neighborhood. We rule out any extended nearest neighbor effects, and deduce the electronic differences are a result of difference in adatom atomic species, i.e., Sn vs S adatoms.. To confirm, density functional theory (DFT) calculations were used to determine the cause of the different electronic behavior and verify our findings.

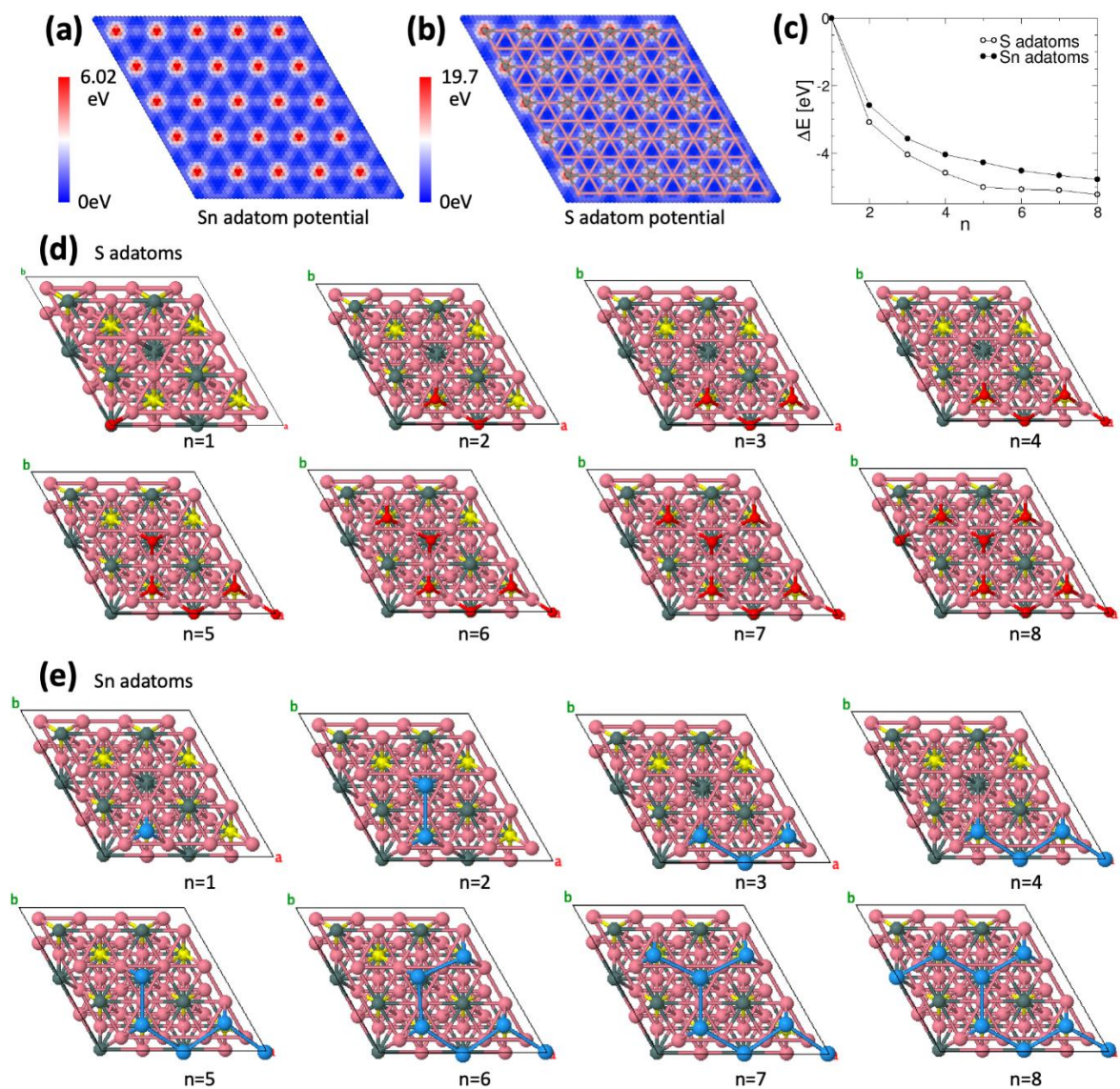

**Figure 7**: First-principles modeling of S and Sn adatoms on $Co_3Sn_2S_2$ surface. Adatom potential of (a) S and (b) Sn on surface. (c) Total energies of S and Sn adatoms as a function of size (*n*) of adatom aggregates. Atomic configurations of most stable aggregates for given *n* for S and Sn are shown in (d) and (e). Pink, grey, and yellow stand for Co, Sn, and S. Red and blue atoms are S and Sn adatoms, respectively, on the surface.

S and Sn atoms were introduced as adsorbates on the surface. Figure 7 shows the potential landscape of S and Sn atoms as adsorbers on the 5x5x1 surface. The 2D potential maps illustrate changes in the adsorption energies of the adatoms as a function of the x- and y-coordinates on the surface (see the underlying lattice structure in Fig. 7 (b)). The energy refers to the most stable location (blue). The potential energy level can be up to 6.02eV for Sn and 19.7eV for S as shown in Fig.7(a) and 7(b), respectively, indicating that Sn can be much more mobile on the surface than S. The total energy per adatom decreases as the adatoms form a denser network, and the lowest energy is attained when the 2D hexagonal lattice of the adatoms fully covers the surface; see the corresponding atomic configurations in Fig. 7(d) and 7(e). The top layer and adatoms are fully relaxed using an optimization algorithm[27] with the atomic force less than 0.01eV/Å.

By excluding the possibility by which we postulate the difference in spectral features among similarly organized adatoms could be a result of second or further nearest neighbors, we are left with the assumption where we surmise the difference is due instead to the physical difference of atomic species. This argument on different adatom species is strengthened bythe theoretical calculations described above, which show that both Sn and S adatoms are energetically feasible on the shandite surfaceEvidently S adatoms are more likely to be present due to lower energy requirements; however, this does not preclude both atoms from being simultaneously present. Additionally, given that the DFT results show that Sn adatoms have increased surface mobility compared to S, they may also be more likely to form larger aggregates of varying configurations, which will have lower overall energies. Therefore, the prediction of how adatoms will organize and relative to their atomic makeup is challenging, which is where the demonstrated ML approach helps provide answers. Although a direct comparison between the calculated projected density of states (PDOS) and the experimental or ML extracted dI/dV is challenging due to tip-sample interactions and other effects unable to be taken into account in simulations, a qualitative comparison can nevertheless still be made, and is portrayed in Figure 8, where we show the PDOS for Sn and S adatoms compared to the LDOS calculated from the PCA classes. Each structural class separated in the energy domain exhibits a splitting of the electronic structure near -0.2 V, while the same occurs for the difference between S and Sn for each configuration. This further suggests  different adatom atomic species can be present in all different configurations. Peculiarly, the height measurements of the expected Sn adatom in Figure 1 do not correspond to the same adatoms separated using the ML methods. We propose that due to the CoSn terminated surface, there exist regions that are topographically lower in height where the Sn atoms – which are more mobile as shown by DFT – can find an energy minimum. Despite this, all other details point to the presence of both S and Sn, and ultimately the changes in electronic structure are revealed due to various configurations.

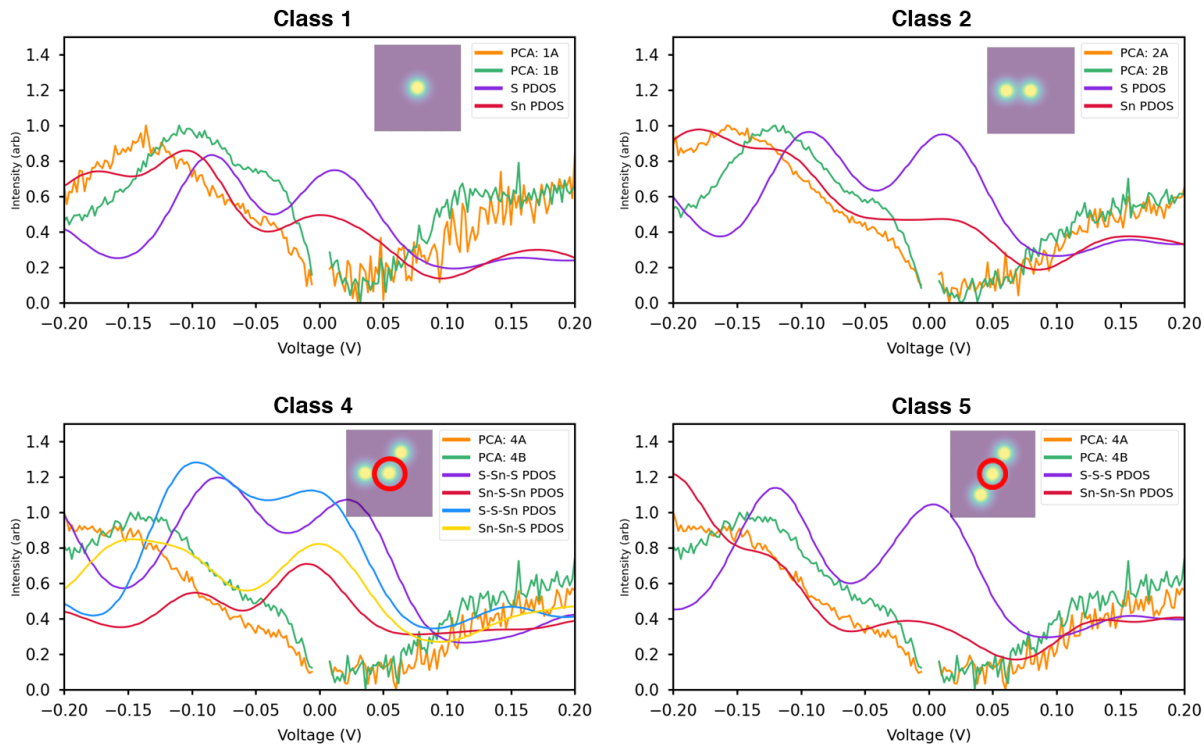

**Figure 8**. Qualitative comparison of calculated PDOS to ML extracted spectra for each class. Vertical axes are in arbitrary units, where PDOS has been scaled to closely match the same vertical scale as experimentally acquired data. Class separation observed in DOS of ML separated spectra is also observed when choosing different adatom atomic species in calculations of PDOS in all classes for various combinations of Sn and S.

Considering the unexpected complexity of adatom distributions, configurations, and electronic properties on the $Co_3Sn_2S_2$ surface, the appropriate choice of ML methods provides a great deal of insight into relevant physics, and importantly provides an exploratory tool for unknown or unexpected phenomenon. We found that structural classification of surface adatoms to obtain a configuration-electronic property correlation is a task well-suited for ML. In the process, we demonstrated the power of using ML to reveal features not easily recognized within such hyperspectral data. We successfully identified additional complexity in the surface of the Type I cleaved $Co_3Sn_2S_2$ surface due to different adatom configurations, atomic species, and consequently variation in electronic properties, all of which affect the surface topology of $Co_3Sn_2S_2$, which we expect will have a significant impact on the future studies of similar surfaces.

The developed analysis workflow can be used universally for the analysis of the electronic properties and structure-property relationships on complex surfaces to explore the emergence of the collective phenomena and chemical interactions. We further note that incorporation of explicit physical knowledge in the analysis, e.g., using the descriptors centers at the atomic sites, incorporation of possible atomic configurations as classes, etc. significantly increases the veracity of ML analysis and offers multiple opportunities at the boundary between physics and ML.

**Acknowledgements:** This effort (feature extraction, machine learning, and first-principles modeling) is based upon work supported by the U.S. Department of Energy (DOE), Office of Science, Basic Energy Sciences (BES), Materials Sciences and Engineering Division (K.M.R., S.V.K., M.Y.). Scanning tunneling microscopy (ZG) and STM simulations were conducted at the Center for Nanophase Materials Sciences (CNMS), a U.S. Department of Energy Office of Science User Facility. This research used resources of the Oak Ridge Leadership Computing Facility and the National Energy Research Scientific Computing Center, DOE Office of Science User Facilities.

**Methods**

**Sample growth and characterization**:

$Co_3Sn_2S_2$ single crystals were synthesized using the self-flux method using a procedure similar to that described by Ref [9]. Co slugs(Alfa Aesar, 99.995%), Sulfur pieces(Alfa Aesar, 99.9995%) and Sn shots(Alfa Aesar, 99.99+%) with an atomic ratio of Co:S:Sn=9:8:83 were placed in a 2 ml Al2O3 Canfield crucible set[10] and sealed into a silica tube under vacuum. The tube was heated to 400°C at 100°C/h. After dwelling for 4 hours, the tube was heated with the same rate to 1100°C and kept at this temperature for 24 hours. The tube was then cooled to 700°C at 3°C/h prior to separating the flux from the crystals in a centrifuge.Magnetic measurements were performed using a Quantum Design magnetic property measurement system (MPMS) which has the reciprocating sample option (RSO) and ac susceptibility option. Phase purity, crystallinity, and the atomic occupancy of the $Co_3Sn_2S_2$ crystals were checked by collecting powder X-ray diffraction (XRD) data. Crystals were cleaved in ultra-high vacuum (UHV) at ~ 78 K and then immediately transferred to the scanning tunneling microscopy/spectroscopy (STM/S) head which was precooled to 4.2 K or 78 K without breaking the vacuum. The STM/S experiments were carried out at 4.2 K or 78 K using an UHV low-temperature, high-field STM with a base pressure better than $2\times10^{-10}$ Torr. W tips were chemically etched then conditioned on clean Au (111) and checked using the topography, surface state and work function of Au (111) before each measurement. The STM/S was controlled by a SPECS Nanonis control system. Topographic images were acquired in constant current mode with bias voltage applied to sample and with the tip grounded. All the spectroscopies were obtained using the lock-in technique with a modulation of 0.1 to 1 mV at 973 Hz on bias voltage, dI/dV.

**Calculations**:

The DFT calculations were performed using Vienna *ab initio* simulation package (VASP)[24] with the projector augmented wave method. We used Perdew-Burke-Ernzerhof (PBE) functional[25] for the exchange-correlation functional. The energy cutoff of the plane-wave basis set is 400 eV, and all atoms are fully relaxed until the residual forces on each atom are smaller than 0.01 eV/Å. The experimental structure of Shandite $Co_3Sn_2S_2$, a rhombohedral lattice structure in the space-group

R3m (No. 166)[26] was used. We then constructed a slab in the (001) crystallographic orientation with CoSn termination, which is consistent with our experimental structure. The Co atoms on the surface form a Kagome lattice with the Sn atom in the center. Our supercell structure contains 4x4x1 unit cells with 512 atoms and a thickness of ~15 Å.

**Supplementary Materials**

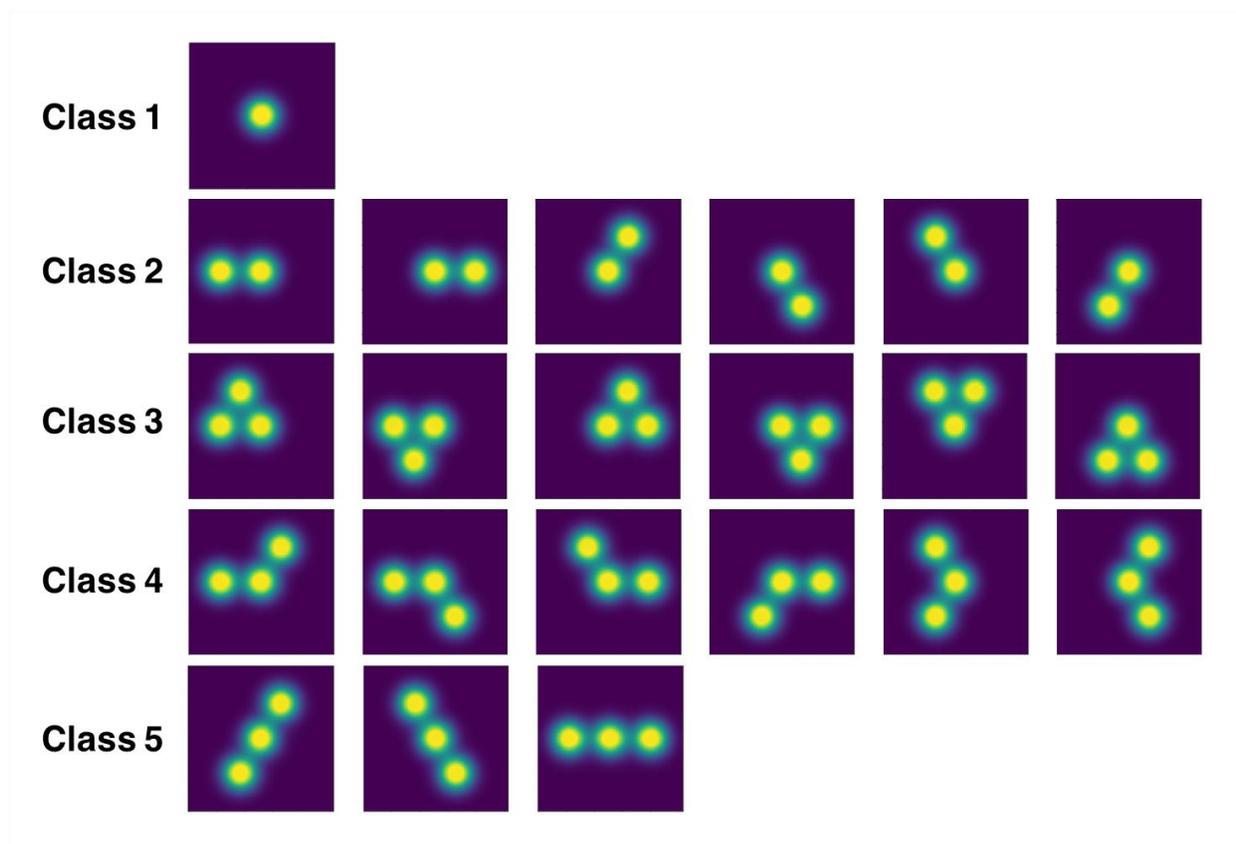

**Figure S1**: Selected distinct classes and their permutations for different adatom nearest neighbor configurations

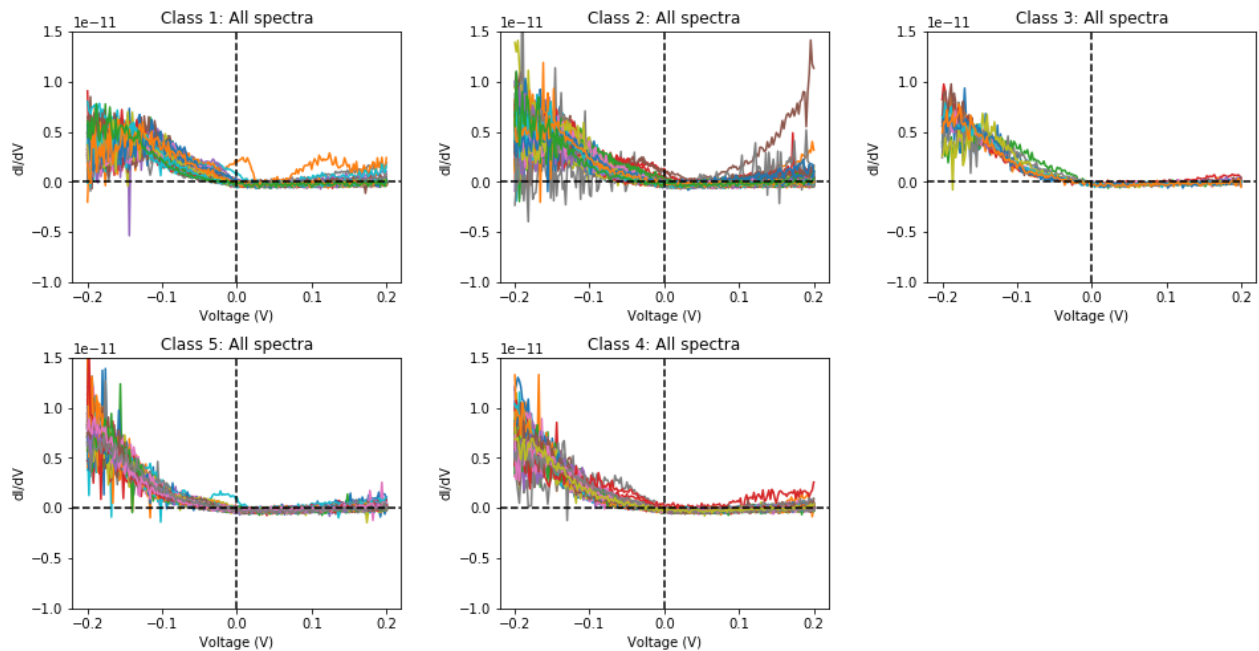

**Figure S2**. All raw spectra of each class. The few spectral outliers seen tend to be due to the probe being far enough off the atom to detect substrate effects.

To test whether or not an extended neighborhood is affecting the spectral response of the adatoms, we performed a Gaussian mixture model on a stack of cropped images containing only class 2 atoms. The window size of the images in this stack was varied to include a larger neighborhood than just the first nearest neighbors. The clustering detected based on the extended structure were almost spectrally identical, as seen in Figure S3, indicating that an extended neighborhood likely is not the reason for the difference in electronic behavior.

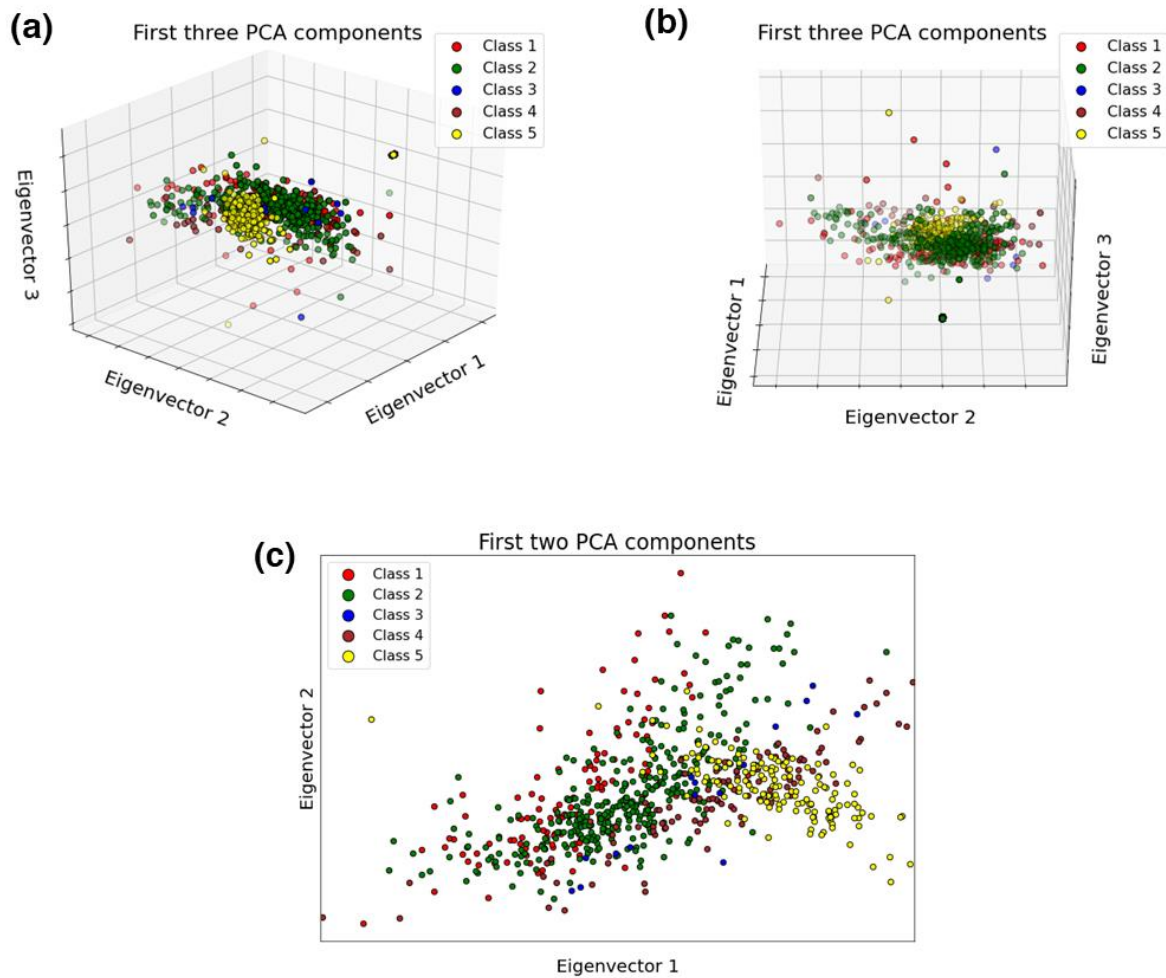

**Figure S3**: Clustering the structural classes in PCA space by GMM. Shown in 3-dimensional PCA space with irst three eigenvectors. (a) and (b) show two different viewing angles in 3-dimensional PCA space, while (c) shows the 2-dimensional clustering visualization in PCA space for the first two eigenvectors. In both two and three dimensional PCA space, Class 2 is seen to spread in PCA space more than the others, suggesting possible additional classification.

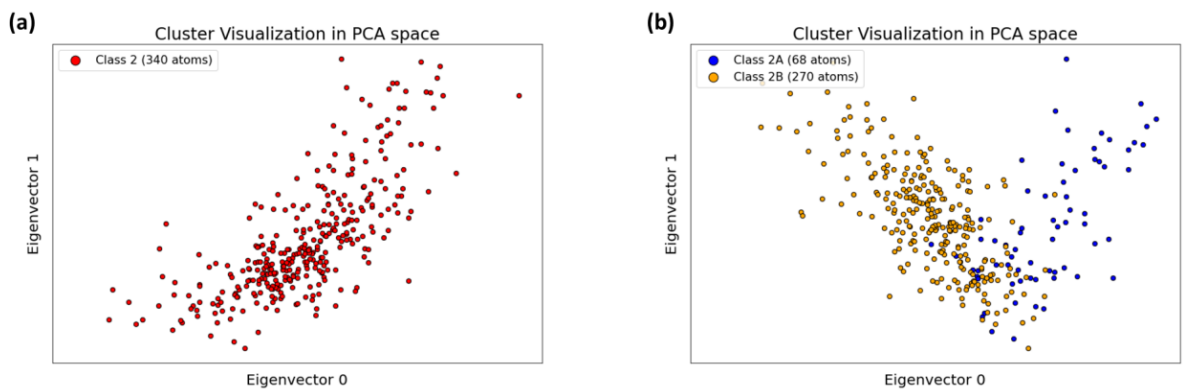

**Figure S4**. Visualization of clustering in 2-dimensional PCA space for class 2 before further spectral classification (a) and after (b). The spectra are decomposed into three components, and the two-dimensional space viewed is that between the first and second components, i.e., eigenvector 0 and 1. Note in (b) there is clear difference in PCA space between class 2A and 2B, implying two distinct classes indeed exist.

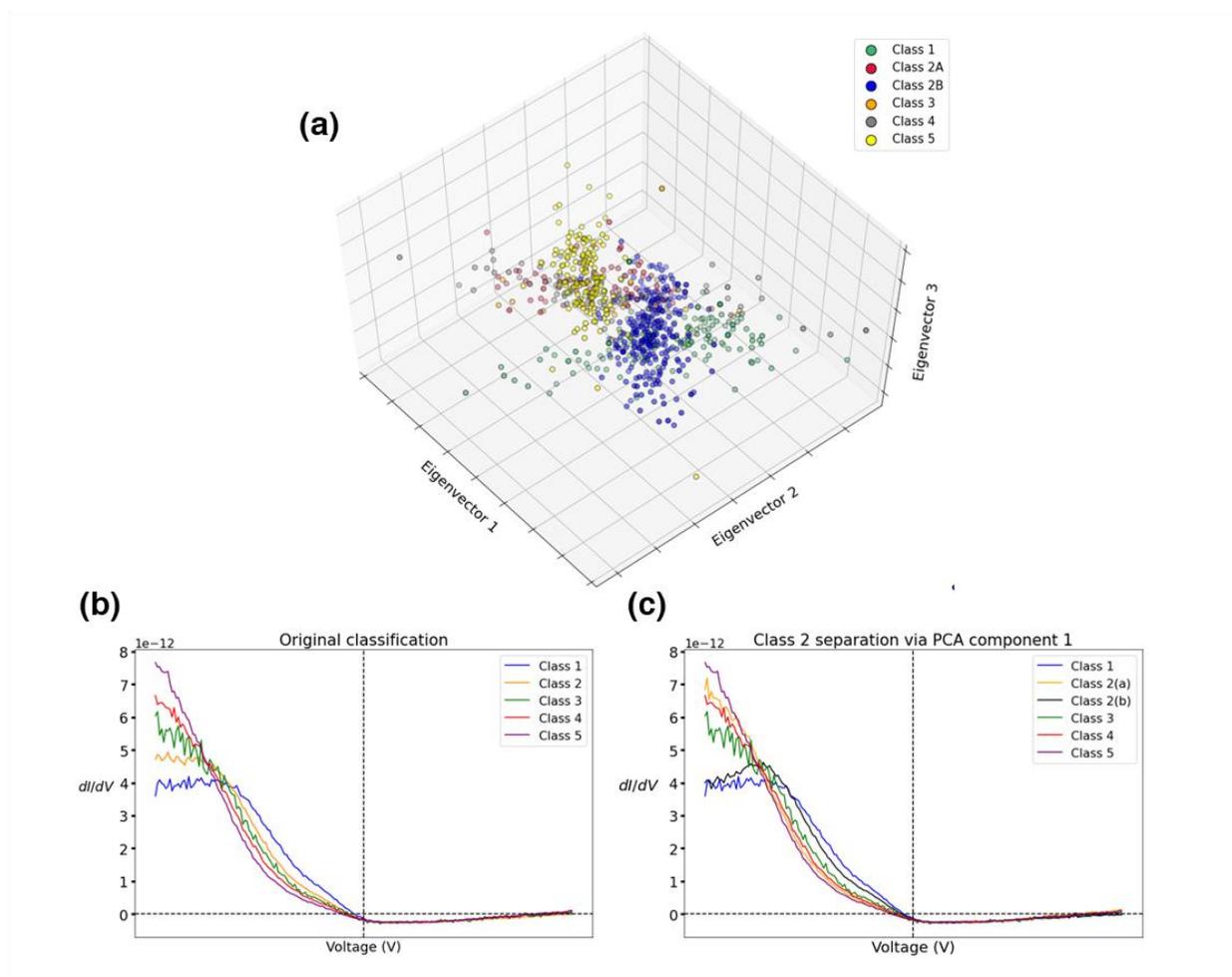

**Figure S5**: Classification refinement outcome. PCA clustering in three dimensions shown in (a), *(b)* shows original classification, while *(c)* is after spectrally deconvolving class 2 atoms into two classes based on energy space.

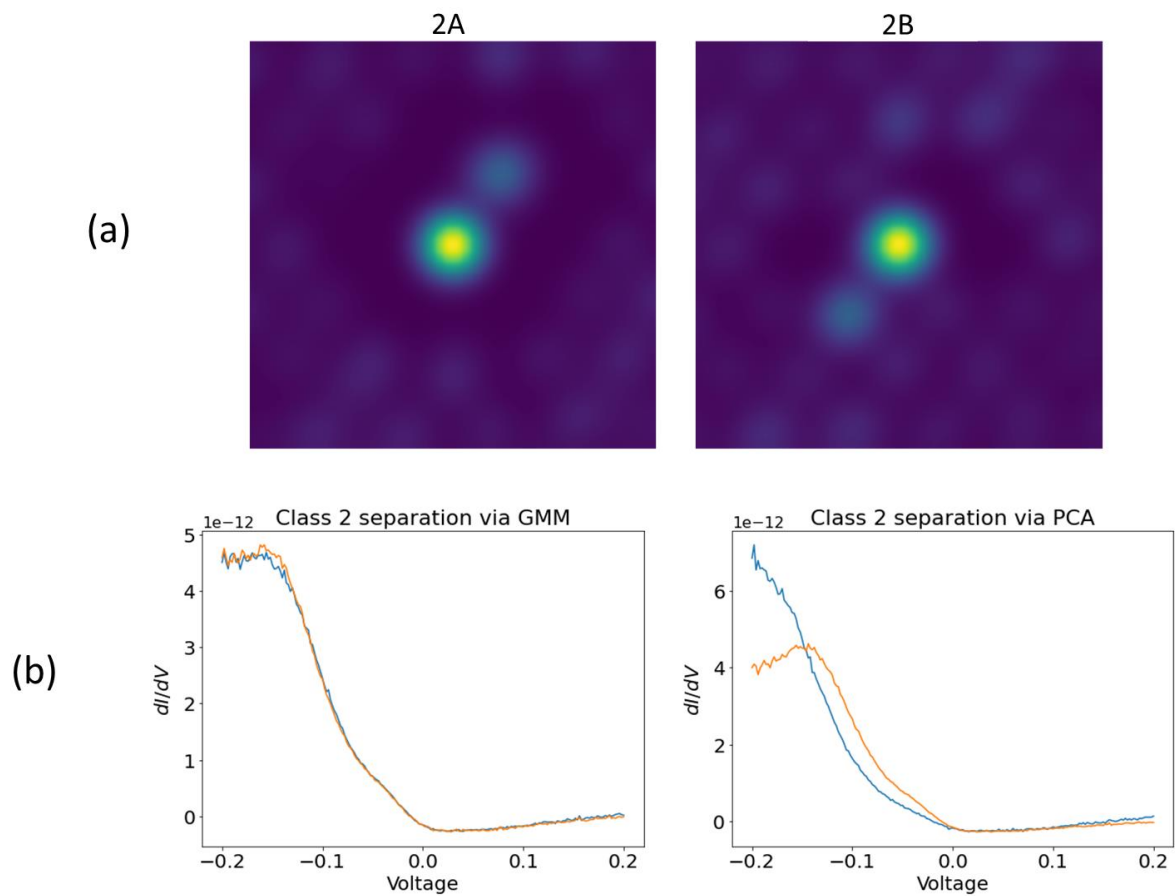

**Figure S6**. Extended neighborhood effect. Attempt to separate class 2 structurally with an extended neighborhood using a Gaussian mixture model (GMM) approach with larger window size to include more neighboring atoms in the analysis. (a) shows the GMM clustering of Class 2 into two subclasses, 2A and 2B, based on topographical information of several unit cells. Panel (b) shows the average spectra of the two GMM subclasses of class 2 (*left*) compared to the two subclasses resulting from spectral decomposition with PCA (*right*).